\documentstyle[12pt,psfig]{article}

\newcommand{\la}{\label}

\newcommand{\AmS}{{\protect\the\textfont2
\renewcommand{\thesection}{\Roman{section}}
  A\kern-.1667em\lower.5ex\hbox{M}\kern-.125emS}}
 
\begin{document}
%\rightline {Report no.}
\vskip 2. truecm
\centerline{\bf Finite Density Fat QCD}
\vskip 2 truecm
\centerline { R. Aloisio$^{a,d}$, V.~Azcoiti$^b$, G. Di Carlo$^c$, 
A. Galante$^b$ and A.F. Grillo$^d$}
\vskip 1 truecm
\centerline {\it $^a$ Dipartimento di Fisica dell'Universit\`a 
dell'Aquila, L'Aquila 67100 (Italy).}
\vskip 0.15 truecm
\centerline {\it $^b$ Departamento de F\'\i sica Te\'orica, Facultad 
de Ciencias, Universidad de Zaragoza,}
\centerline {\it 50009 Zaragoza (Spain).}
\vskip 0.15 truecm
\centerline {\it $^c$ Istituto Nazionale di Fisica Nucleare, 
Laboratori Nazionali di Frascati,}
\centerline {\it P.O.B. 13 - Frascati 00044 (Italy). }
\vskip 0.15 truecm
\centerline {\it $^d$ Istituto Nazionale di Fisica Nucleare, 
Laboratori Nazionali del Gran Sasso,}
\centerline {\it Assergi (L'Aquila) 67010 (Italy). }
\vskip 3 truecm

\centerline {ABSTRACT}
\vskip 0.5truecm

\noindent
Lattice formulation of Finite Baryon Density QCD is problematic
from computer simulation point of view; it is well known that for
light quark masses the reconstructed partition function fails to
be positive in a wide region of parameter space. 
For large bare quark masses, instead, it is possible 
to obtain more sensible results; problems are still present but 
restricted to a small region.
We present evidence for a saturation transition independent from
the gauge coupling $\beta$
and for a transition line that, starting from the 
temperature critical point at $\mu=0$, moves towards smaller $\beta$
with increasing $\mu$ as expected from simplified 
phenomenological arguments.
\vfill\eject

\section{Introduction}

From the point of view of computer simulation, lattice approach to
non perturbative aspects of quantum field theory is a mature technique; 
apart from few exceptions, well consolidated schemes of 
simulation do exist, something like a recipes book, that allow studies, for 
example, of the most interesting features of QCD.
The progress in the results is quite slow, in view of the large 
computing power needed for realistic calculation, but the field 
appears well founded. 

The mentioned "few exceptions", however, concern very interesting 
problems, as well. The most paradigmatic of these dark zones
is the study of thermodynamic of QCD in presence of non-zero 
baryonic density, shortly Finite Density QCD.
The standard way to include the effects of baryonic matter on QCD 
vacuum leads to complex action in Euclidean formulation and this
prevents the use of standard simulation algorithms, based on the 
idea of importance sampling, defined through a positive definite 
density of probability, {\it e.g.} the exponential of minus the Euclidean 
action. 
This problem can be rephrased stating the impossibility of defining a 
Boltzmann weight for 
each field configuration: only calculating the partition function we can
define correctly the observables and obtain sensible results for quantities 
of physical interest.

Calculations of partition functions are not infrequent in lattice 
simulations \cite{bhakar}, but their nature of extensive quantities raises the
problem of the feasibility of this type of calculation with 
limited statistics, as forced from finite computing power.
In the following section we will argue that, although reliable evaluation
of the partition function of fermions coupled lattice gauge theories at
zero baryon density is possible and successful \cite{azc}, 
the extension of such technique for Finite Density QCD appears 
out of reaching for any reasonable statistics, at least in a range of theory 
parameters: for some values of the chemical potential $\mu$ 
the phase of the fermionic determinant 
can be estimated only averaging over $O(e^V)$ configurations. 

In two recent papers \cite{noi1,noi2} we have presented our results for Finite 
Density QCD, mainly working with light quarks, in the infinitely strong
coupling limit. Our approach consisted in trying to approximate  the correct 
partition function using the modulus of the determinant, as suggested
by the analysis of the SU(3) linear chain model \cite{bilic}, \cite{lat97}.
The emerging picture  is quite disappointing: we are
essentially unable to reproduce the results of ref. \cite{karsh}, obtained 
with the MDP (Monomer-Dimer-Polymer) technique.
Moreover we did show as the simulation scheme,
known as Glasgow method \cite{bar}, and possibly any other method based 
on the Grand Canonical Partition Function (GCPF) approach \cite{gibbs}, 
produce essentially the same results of
our calculations, when analyzed in a way to avoid perverse numerical
effects due to rounding errors \cite{noi2}. 

In spite of these disappointing aspects, there are regions in the 
parameters space of the theory, in particular at large  bare quark masses, 
in which one can hope to obtain reliable
results, of some interest from a methodological point of view. 
In fact, at large quark masses (and any $\beta$), 
the interval of chemical potential where the contribution of the phase can
not be appreciated shrinks. 

The rest of the work, consequently,  is devoted to the investigation of 
large quark mass limit, in some sense a favorite laboratory in which 
numerical techniques can be tested.
Monitoring the expectation value of the phase of the Dirac determinant
we can distinguish the regions in the parameter space where our evaluation
of the partition function of finite density QCD is (in principle) exact 
from the ones where we miss a possible contribution to ${\cal Z}$.
A coherent picture seems to emerge from our data: 
a saturation transition exists at all couplings
and merges, in the scaling region, to the true deconfining critical line
that, with respect to $\mu=0$, moves towards smaller 
$\beta$ with increasing $\mu$.

In the next section we will give arguments to explain why the contribution
of the phase can not be measured and will present, 
in the strong coupling limit,
a quantitative check of the Grand Canonical formulation results using
data obtained with different techniques. 
The third section is devoted to the exposition of our approach to 
simulations of finite density QCD at finite coupling which  exploits the 
main advantage of the MFA approach, {\it i.e.} the free
mobility in the $(\beta, \mu)$ plane. In the fourth section we 
present results for fermionic and gluonic observables, discussing 
the fate of the deconfining phase transition, expected, on phenomenological 
ground, when one increases the baryon density. 
The analisys is complemented with informations coming from the 
partially solved infinite bare mass limit \cite{noi3}.
In the last
section a final discussion of the most important results is done.

\section{The partition function of Finite Density QCD}

The Finite Density QCD partition function can be written as

\begin{equation}\la{z}
{\cal Z} = \int [dU] e^{-\beta S_g(U)} \det\Delta(U,m_q,\mu)
\end{equation}

\noindent
where, using the staggered formulation, 
the fermionic matrix $\Delta$ takes the standard form \cite{hase}

\begin{eqnarray}
\Delta_{i,j}=m_q\delta_{i,j}+\frac{1}{2}\sum_{\nu=1,2,3}\eta_{\nu}(i)
[U_{\nu}(i)\delta_{j,i+\hat{\nu}}-
U_{\nu}^{\dag}(i-\hat{\nu})\delta_{j,i-\hat{\nu}}]\nonumber \\
+\frac{1}{2}[U_4(i)\delta_{j,i+\hat{4}}e^{\mu}-
U_4^{\dag}(i-\hat{4})\delta_{j,i-\hat{4}}e^{-\mu}]\nonumber
\end{eqnarray}

\noindent
The contribution of modulus $|\det\Delta |$ of Dirac determinant and
its phase $\phi_\Delta$ can be separated as

\begin{equation}\la{zz}
{\cal Z}={\cal Z}_\|\langle e^{i \phi_\Delta}\rangle_\|
\end{equation}

\noindent
where

\begin{equation}\la{z||}
{\cal Z}_\| = \int [dU]e^{-\beta S_g(U)} |\det\Delta(U,m_q,\mu)|
\end{equation}

\noindent
is the partition function of the model with the modulus of the determinant
(modulus QCD in the following), and

\begin{equation}\la{fas}
\langle e^{i \phi_\Delta} \rangle_\| =
\frac
{\int [dU] e^{-\beta S_G(U)} |\det\Delta | e^{i\phi} }
{\int [dU] e^{-\beta S_G(U)} |\det\Delta | }
\end{equation}

It is clear from eq. (\ref{zz}) that, in the thermodynamical limit,
the theory defined by means of ${\cal Z}_\|$ is physically different
from the original theory only when the expectation value of the cosine of
the phase of fermion determinant is vanishing exponentially with the system 
volume. In the regions of parameter space where the
aforementioned expectation value is not $O(e^{-V})$ 
modulus QCD is an equivalent formulation of Finite Density QCD {\it i.e.}
indistinguishable in the thermodynamical limit. 
In the rest of parameter space, modulus QCD clearly  
overestimates the true QCD partition function.

Let us try to better illustrate this concept looking at figure 1. It refers
to infinitely strong coupling limit $\beta=0$ and $V=6^3\times 4$. 
At fixed quark mass $m_q$
the partition function of the system is only dependent on chemical potential
$\mu$. If we plot the free energy versus $\mu$ we can extract the
phase structure from the appearance of a singularity in (some derivative of)
the curve.

Two extreme limits are well known. At $\mu=0$ we get the logarithm of
the usual fermion determinant averaged over gauge field configurations
with a flat distribution: an average of a well defined (real and positive)
quantity that can be computed.
On the other hand in the large $\mu$ limit only the last term 
of Grand Canonical Partition Function (see later) survives:
$\det \Delta \to \left (1/ 2\right )^{3V}e^{3V\mu}$ and 
the free energy is a straight line with slope $3V$. 
In this limit the (baryon) number density, defined as:

\begin{equation}\la{nmu}
N(\mu)=\frac{1}{3V} \frac{\partial}{\partial\mu} \log{\cal Z}
\end{equation}

\noindent
is equal to $1$, and we can say that we are in a saturation regime, with
Pauli exclusion principle preventing further increase of baryon density.
In these two limits modulus QCD is coincident with the true
theory and deviations are possible only in the intermediate region.

Starting from $\mu=0$, we can use the data of fig. 5 in ref. \cite{karsh},
regarding number density at $m_q=0.1$, in order to reconstruct
the free energy of the true theory as seen from the MDP approach.
This is shown in figure 1 as the dotted line. If we superimpose the results
of modulus QCD (continuous line) we can easily identify three regions:

\begin{itemize}
\item {$\mu<\mu_1=0.3$, which defines the onset in modulus QCD, where the 
number density is essentially zero;}
\item {$\mu>\mu_2=1.0$ , the saturated region;}
\item {$\mu_1<\mu<\mu_2$, the region where modulus QCD grossly overestimates
the free energy of true theory.}
\end{itemize}

\noindent
(as stated in \cite{noi2}, using Glasgow prescription for dealing with the 
complex determinant, we obtain, for the free energy, exactly the
same results as in modulus QCD).

In figure 2 we report, for the same lattice and quark mass, 
the difference between the free energy of modulus QCD 
and the estimation based on data of ref. \cite{karsh}.
Superimposed to that we plot the expectation value of 
$\langle e^{i \phi_\Delta}\rangle_\|$ at the same
value of the parameters. It is evident that the intermediate region is
where the phase term is vanishing within statistical
errors. If we concentrate on a value of $\mu$ inside this region,
for example $\mu=0.7$, and we plot the distributions of the phase and the 
(logarithm of the) modulus of fermion determinant of 
single field configurations,
we can see (figure 3) that modulus distribution is behaved as expected,
while the phase distribution is almost flat. 

These distributions have been
computed using $N\simeq 2500$ configurations of a $6^3\cdot 4$ lattice.
With this statistics we can
hope to measure accurately the phase term $\langle e^{i \phi_\Delta}\rangle_\|$
only down to $O(1/\sqrt{N})$ ($\simeq 0.02$ for our runs), 
far from the $O(e^{-V})$ order needed in principle. 
Even with a statistics of
some thousands of configurations, we can say nothing on free energy of
true theory in the range $\mu_1<\mu<\mu_2$,
that covers the region where the number density varies rapidly.
This does not imply necessarily that the phase is relevant in this region: 
for example it could go to zero as $e^{-V_S}$, with $V_S$ the spatial volume, 
being in this case at the same time irrelevant
and non measurable!

The situation becomes somewhat better if we move to large quark mass:
the range $(\mu_1,\mu_2)$, where finite statistics effects prevent to 
obtain a sensible evaluation of free energy, becomes narrower (see later), 
thus allowing the study of the model in a wider parameters range. 
The same scenario holds at finite coupling too, allowing us to investigate
a great part of the parameter space.

\section{Simulation Scheme}

In this section we will present the simulation scheme that we have used in 
our work. 
Our simulations are based on the GCPF (Gran Canonical Partition Function) 
formalism with an MFA (Microcanonical Fermionic Average) inspired approach 
for intermediate coupling analysis. 

The GCPF formalism allows one to write the fermionic determinant as a 
polynomial in the fugacity $z=e^{\mu}$:

\begin{eqnarray}\la{det}
\det\Delta(U;m,\mu)&=&\det(G+e^{\mu}T+e^{-\mu}T^{\dag})=
z^{3V}\det(P(U;m)-z^{-1})\nonumber \\
&=&\sum_{n=-3L_s^3}^{3L_s^3}a_nz^{nL_t}\nonumber
\end{eqnarray}

\noindent
where the propagator matrix $P$ \cite{gibbs} is

\[ P(U;m) = \left( \begin{array}{cc}
-GT & T \\
 -T & 0  \end{array} \right) \\
\]

\noindent
in which $G$ contains the spatial links and the mass term, $T$ contains the 
forward temporal links and $V=L_s^3L_t$ is the lattice volume.

Once fixed the quark mass $m_q$, a complete diagonalization of $P$ allows one
to reconstruct, trough a recursion algorithm, the coefficients $a_n$, hence
$\det\Delta$, for any value of the chemical potential $\mu$.
Due to the $Z(L_t)$ symmetry of the eigenvalues of $P$ it is possible to write
$P^{L_t}$ in a block form and we only need to diagonalize a 
$6L_s^3\times6L_s^3$ matrix.

This general method has been implemented in the framework of an MFA \cite{mfa}
inspired approach.
The basic idea in MFA is the exploitation of the physical equivalence between
the canonical and microcanonical formalism via the introduction of
an explicit dependence on the pure gauge energy in the computation of the 
partition function. Indeed (\ref{z}) can be written as:

\begin{equation}\la{zmfa}
{\cal Z}(\beta,\mu,m)=\int dE n(E) e^{-6V\beta E}
<S_{eff}^F(\mu,m_q)>_E
\end{equation}

\noindent
where 

\begin{equation}\la{ne}
n(E)=\int [dU] \delta(6VE-S_g[U])
\end{equation}

\noindent
is the density of states at fixed pure gauge energy $E$, and

\begin{equation}\la{az}
<S_{eff}^F(\mu,m_q)>_E=
\frac{\int [dU] \delta(6VE-S_g[U])S_{eff}^F([U],\mu,m_q)}
{n(E)}
\end{equation}

\noindent
is the average over gauge field 
configurations at fixed energy $E$ of a suitable definition of effective
fermionic action.

For the calculation of $<S_{eff}^F>_E$ we proceed
as follows: first, we choose a set of energies selected 
to cover the range of $\beta$ we are interested in.
Secondly, for all the energies in the set, we generate gauge field 
configurations using a pseudo-microcanonical code; the generation 
of gauge fields at fixed energy is not the costly part of the whole procedure, 
so we can well decorrelate the configurations used for measuring the Dirac 
operator. Then, a standard NAG routine is used in order to obtain the complete 
set of eigenvalues of the propagator matrix $P$.
At this point we can reconstruct the fugacity expansion coefficients $a_n$
or, without any substantial additional computer cost,
use the eigenvalues to explore the possibilities offered by
alternative prescriptions for the fermionic effective action, {\it i.e.}
evaluate the modulus of the determinant and hence ${\cal Z}_\|$.
At the end, we have the fermionic effective action
evaluated at discrete energy values: a
polynomial interpolation allows the reconstruction at arbitrary values of
the energy $E$, in order to perform the numerical one-dimensional
integration in (\ref{zmfa}) 
and obtain the partition function ${\cal Z}_\|(\beta,\mu,m)$.

In a previous work \cite{noi2} we have found evidence for numerical 
instabilities in the standard evaluation of coefficients $a_n$, 
whose origin lies  on the ordering of the eigenvalues of $P$ 
as calculated by a standard diagonalization routine.
A random eigenvalue arrangement, 
before the calculation of the coefficients $a_n$, is necessary in order to 
control rounding effects.
In the present work we have always used this procedure to calculate the GCPF 
expansion coefficient.

To conclude this section let us, briefly, summarize the usefulness of MFA.
This algorithm does not require a separate fermionic simulation for each 
value of $\beta$, as the standard HMC (Hybrid Monte Carlo) algorithms, thus
allowing us to extend the analysis to the whole relevant $\beta$ range 
without an additional computer cost.

Moreover, the basic idea of MFA is to consider the fermionic determinant 
(or its absolute value) as an observable. So $\det\Delta$ is not in the 
integration measure and one avoids, in principle, the problem of dealing
with a complex quantity in the generation of configurations.
We have seen, however, that an unaffordable (eventually exponential)
statistics is necessary to calculate ${\cal Z}$
in some $\mu$ range so that this advantage on direct simulation schemes 
can not, in general, be exploited.

\section{Large quark mass results}

In this section we will present results in the large  bare quark mass 
limit both in the strong and intermediate coupling QCD at finite density. 
We have performed simulations in a  
$4^3\times 4$ lattice ($10$ masses $m_q=1.0 \to 5.0$)
in the range of the chemical potential $\mu\in[0.0,4.0]$ and, for intermediate
coupling, of $\beta\in[4.0,6.0]$. Some (low statistic) data for a 
$6^3\times 4$ lattice will also be shown.

Firstly we have located the range $(\mu_1,\mu_2)$ (as a function of $\beta$
and $m_q$), in which the average (\ref{fas}) is, with available statistics,
indistinguishable from zero (see fig. 2).
In figure 4 we report the gap $\Delta\mu=\mu_2-\mu_1$ versus the quark 
mass computed fixing the gauge coupling value $\beta=5.5$, a value  
near to the temperature induced transition.
From figure 4 it is evident the tendency of $\Delta\mu$ to decrease as 
the quark mass $m_q$ is increased. 

We have calculated the partition
function of QCD in the $\beta-\mu$ plane. This calculation is in principle
exact for $\mu\notin [\mu_1(\beta),\mu_2(\beta)]$ and is 
complemented with data of modulus QCD in the region where we are not able 
to measure the phase. All the data are presented in such a way to 
make evident when a possible contribution from the phase has been discarded.

In figure 5-a we report $\partial N(\mu)/\partial\mu$
evaluated at $m_q=1.5$ and different values of the gauge coupling 
$\beta$. We have chosen two values of $\beta$: 
$\beta=5.3$ for which the system is inside the confining phase, 
and $\beta=5.7$ in the deconfined one.

The figure shows a sharp peak for the derivative, 
moreover the position of the peak does not move with $\beta$.
The same quantity for the larger lattice is reported in figure 5-b.
We do not attempt to do a finite size scaling analisys but only note that
the height of the peaks grows considerably suggesting a (saturation) 
transition at $\mu=\mu_c^S$ and independent from $\beta$.
The same scenario holds at smaller $m_q$ with the only difference that
the peaks becomes broader for masses smaller than $\simeq 1.0$ signalling
the well known phenomena of early onset in the number density 
\cite{noi1}, \cite{bar}.

These findings are to be compared with the one obtained analytically 
at infinite $m_q$  for which the infinite bare mass 
QCD partition function factorizes \cite{noi3}

\begin{equation}\la{zzz}
{\cal Z}(\beta,\mu) = {\cal R}(\beta,\mu) \cdot {\cal Z}_{PG}(\beta) 
\cdot {\cal Z}(\beta=0,\mu)
\end{equation}

\noindent
with ${\cal R}$ a irrelevant factor in the zero temperature thermodynamical 
limit (${\cal R}\to 1$ for $V=L^4\to\infty$) and ${\cal Z}(\beta=0,\mu)$
developing a first order saturation transition at $T=0$.

Due to the independence of $\mu_c^S$ from $\beta$, we will investigate 
its dependence on the bare quark mass from the
behavior of $N(\mu)$ at $\beta=0$, where we are able to
compare with other results (analytical as well as numerical).
In figure 6 we report $\mu_c^S(m_q)$:
at $m_q \leq 2$  we approach a linear dependence for $\mu^S_c$
in good agreement with the  numerical results by MDP  \cite{karsh}
while, at larger masses, $\mu_c^S$ coincide asymptotically with the
large mass limit $\log(2m_q)$ \cite{noi3} (as well as with the $1/3$
of the nucleon mass \cite{ks}).  

Coming to the phase structure in the $\beta-\mu$ plane it is evident 
that the critical line at constant $\mu=\mu^S_c$ can not be the only one. 
In order to separate the confined phase from the deconfined one we need a
transition line that, starting at the zero density finite temperature
critical point, eventually merges with the saturation transition at
smaller values of $\beta$.
This critical line is the relevant one from a physical point of view
in the sense that here we can obtain finite quantities for physical
observables when the lattice spacing goes to zero.
In the infinite mass limit this line is vertical (at the critical coupling
of the finite temperature pure gauge theory, see formula (\ref{zzz}))
so we can expect that, for our large masses, it moves only slightly to
smaller $\beta$. If this is the case it is crucial the possibility to
rely on data at fixed $\mu$ and any $\beta$ to extract relevant 
informations.

The saturation of number density is a pure lattice artifact 
(a saturated lattice corresponds, in the continuum limit, to a
divergent number density).
Therefore, in order to search for evidence of the transition expected on
phenomenological grounds, it is necessary to restrict our analisys
to values of $\mu$ where the discretization effects are smaller {\it i.e.}
to a number density smaller than the lattice half 
filling value $1/2$.
In this region of the parameter space the phase gives no contribution
(see fig. 2) and we can rely on our results. 

We have studied the plaquette energy 
$E(\beta,\mu)$, the Polyakov loop $P(\beta,\mu)$ and the number density
as a function of $\beta$.

In figure 7-a,7-b we 
report $E(\beta,\mu)$ and $\partial E(\beta,\mu)/\partial\beta$, 
evaluated at bare quark mass $m_q=1.8$ and at different values of the 
chemical potential $\mu<\mu_c^S$. 
In fig. 7-a we can clearly see a rapid variation of the observable for
all the values of $\mu$; 
for the $\mu=0$ curve this happens in correspondence with
the pseudo-temperature transition of zero density full QCD. 
The critical gauge coupling
moves to smaller $\beta$ as we increase $\mu$. This phenomenon is
also evident as a sharp peak in the figure of the derivative (fig 7-b).
It is tempting to interpret this as an evidence of a  temperature induced 
phase transition extending at non zero values of $\mu$.
Also the behavior of the 
Polyakov loop points in this direction: as can be
seen in figure 8, $P_{\mu}(\beta)$ changes rapidly at values of $\beta$ 
consistent with the ones obtained from the energy.
The number density gives a less clear signal since it is forced to be a
constant function of $\beta$ at $\mu=0$.
Nevertheless we can see in figure 9 that the plot of this observable 
is still consistent with previous findings for the gluonic quantities.
It is useful to remark that plotting the same quantities at fixed $\beta$
as a function of $\mu$ we would be practically unable to see any signal.

Signals for a developing discontinuity in all these observables
rely on data in the region where the contribution of the 
phase is negligible but similar behavior is found at larger 
values of $\mu$ too 
(where we can only compute the observables of modulus QCD).
All these findings are consistent with phenomenological expectations 
for the temperature-density QCD phase diagram where, 
increasing the baryonic density, the critical temperature of the 
deconfinement phase transition decreases. 
On the lattice this translates in a critical line that, 
starting at the zero density-finite temperature critical point, 
continues at smaller values of $\beta$ for finite $\mu$.

To conclude our analysis we report in figure 10 the $(\beta,\mu)$ phase diagram
of the theory at $m_q=1.8$ and $V=4^4$.
We can see two (critical) lines; the horizontal one is due to saturation, while
the other should be the physical one. 
If the same scenario holds in the zero mass limit we can expect that,
as the lattice spacing goes to zero,
the latter critical line extends in all the scaling window eventually
coinciding with the former in the zero temperature limit.

\section{Conclusions}

In this paper we have studied Finite Density Lattice QCD by means of numerical 
simulations. As well known this approach, probably the only one
able to tackle the non perturbative effects leading to quark-gluon
plasma transition, suffers severe problems due to the lack of hermiticity
of Dirac operator for a single realization of gauge fields.

In the first part of the paper we have shown as, for small quark masses and
strong coupling, any numerical algorithm based on the GCPF approach gives 
results different from what expected in the region where the contribution
of the phase can not be evaluated.
To our understanding only a statistic exponentially large with the system
volume (and a consequently high accuracy in numerical calculations) 
can solve this problem.

Moving to large quark mass region we meet a much better situation and
a large part of the parameter space becomes accessible to numerical 
simulations.
We get, independently from the gauge coupling $\beta$, a saturation transition 
at a chemical potential $\mu_c^S$ well compatible with the 
one predicted, in the strong coupling regime, by previous numerical 
and analytical analysis. 
The new result is the evidence of another transition line that connects
the previous one to the second order critical point of the four
flavor $\mu=0$ theory.
This has to be regarded as the lattice counterpart of the transition line
in the temperature-chemical potential plane that should separate the
standard hadronic phase from the quark-gluon plasma phase.

For the first time we have got some evidences that the
behavior of finite density lattice QCD can be consistent with standard
phenomenological expectations. 
Larger lattices could clarify the nature of the transitions but the 
volumes attainable with reasonable computer resources
make this program not effective.
To extend these results to the small quark mass region is impossible since
the contribution of the phase to the partition function
becomes not measurable practically in the whole parameter space.
At the end we have to conclude that, until now, finite density lattice 
QCD, far from providing non perturbative quantitative insights in the 
behavior of quarks and gluons, can at most give us some qualitative
indication.

\vskip 0.2 truecm
This work has been partly supported through a CICYT (Spain) - INFN (Italy)
collaboration. 
A.G. was supported by a Istituto Nazionale di Fisica Nucleare fellowship
at the University of Zaragoza.

\newpage
\centerline{\bf Figure Captions}
\vskip 1 truecm
\begin{itemize}

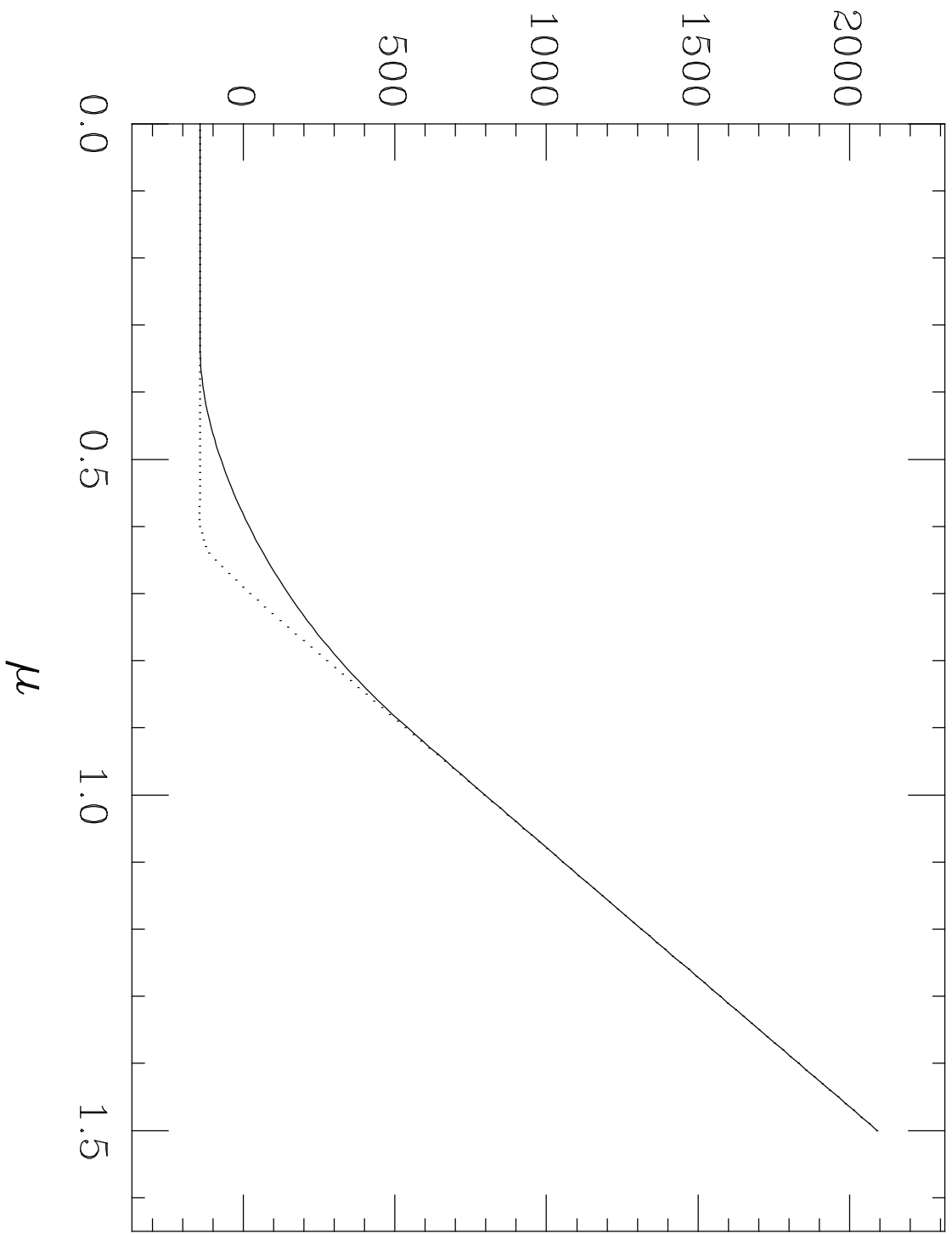
\item{Fig. 1: Free energy at $\beta=0$ and $m_q=0.1$ of true QCD in the 
MDP approach (dotted line) and of Modulus QCD (continuous line).}

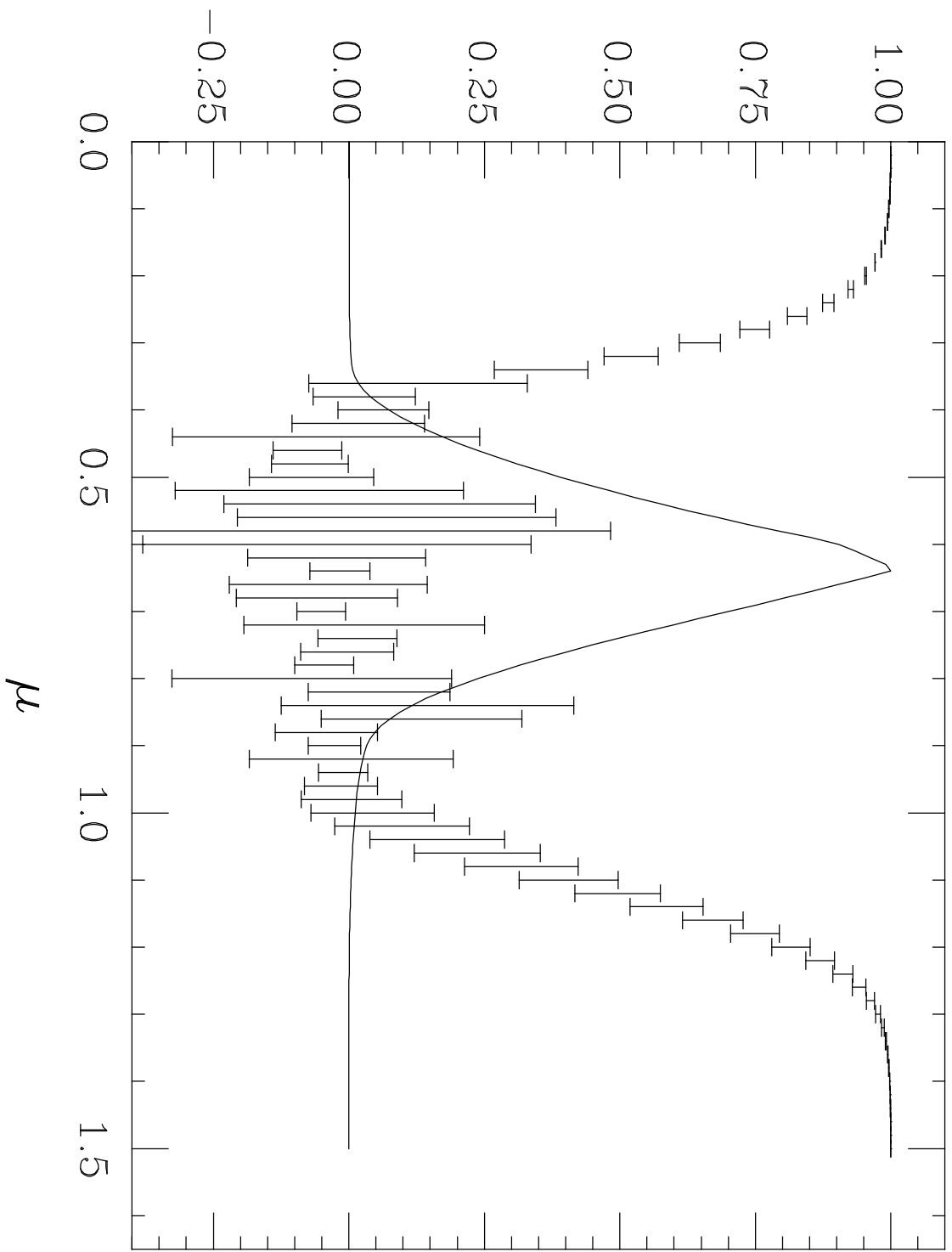
\item{Fig. 2: Normalized difference 
between the free energy of Modulus QCD and the
free energy of true QCD in the MDP approach (continuous line) superimposed
to the expectation value of the determinant phase  
at $\beta=0$ and $m_q=0.1$.}

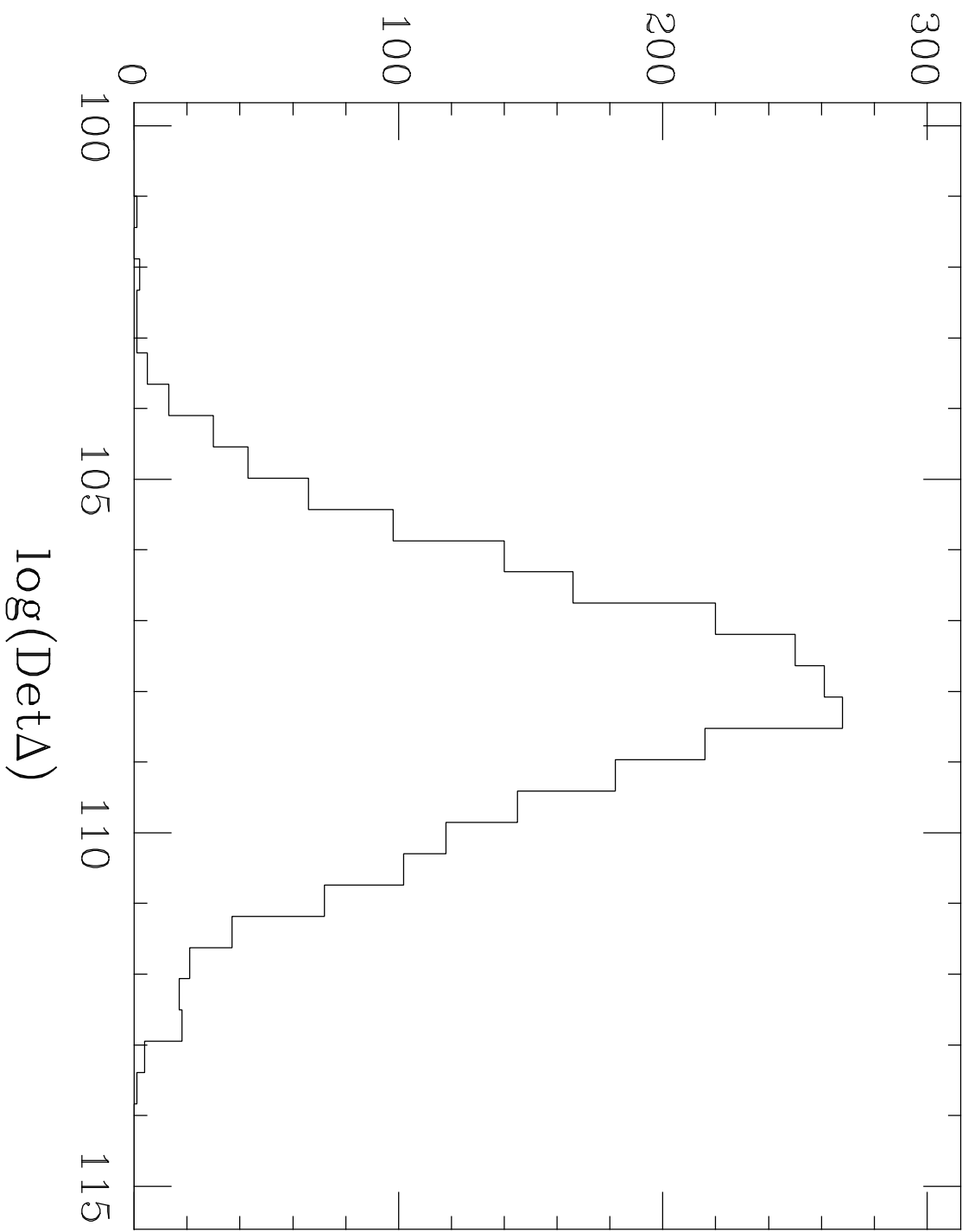
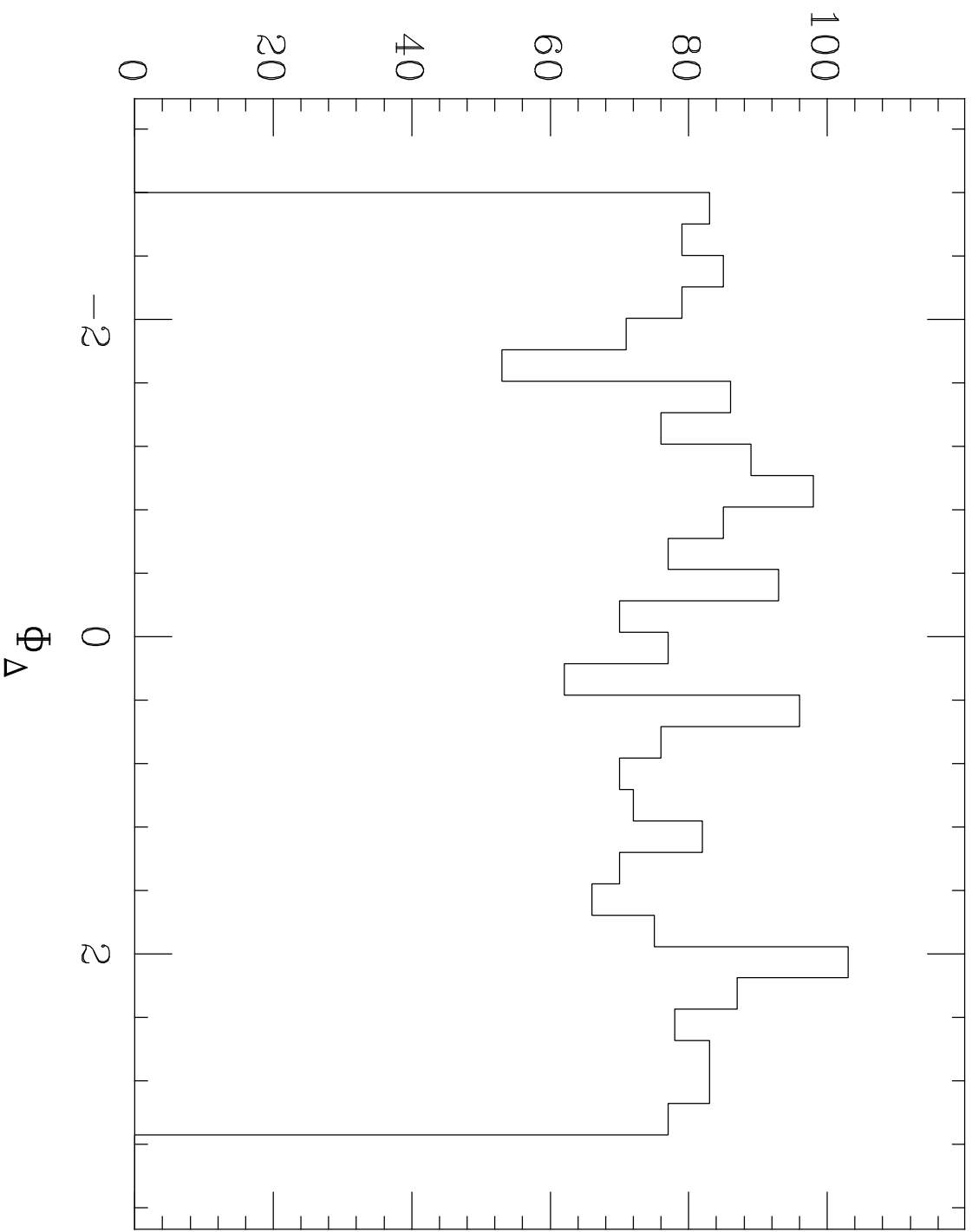
\item{Fig. 3: Distributions of the logarithm of the modulus (a) and of the
phase (b)
of the fermionic determinant in a $6^3\times 4$ 
lattice at $\beta=0$, $m_q=0.1$ 
and $\mu=0.7$ with $N=2500$ configurations.}

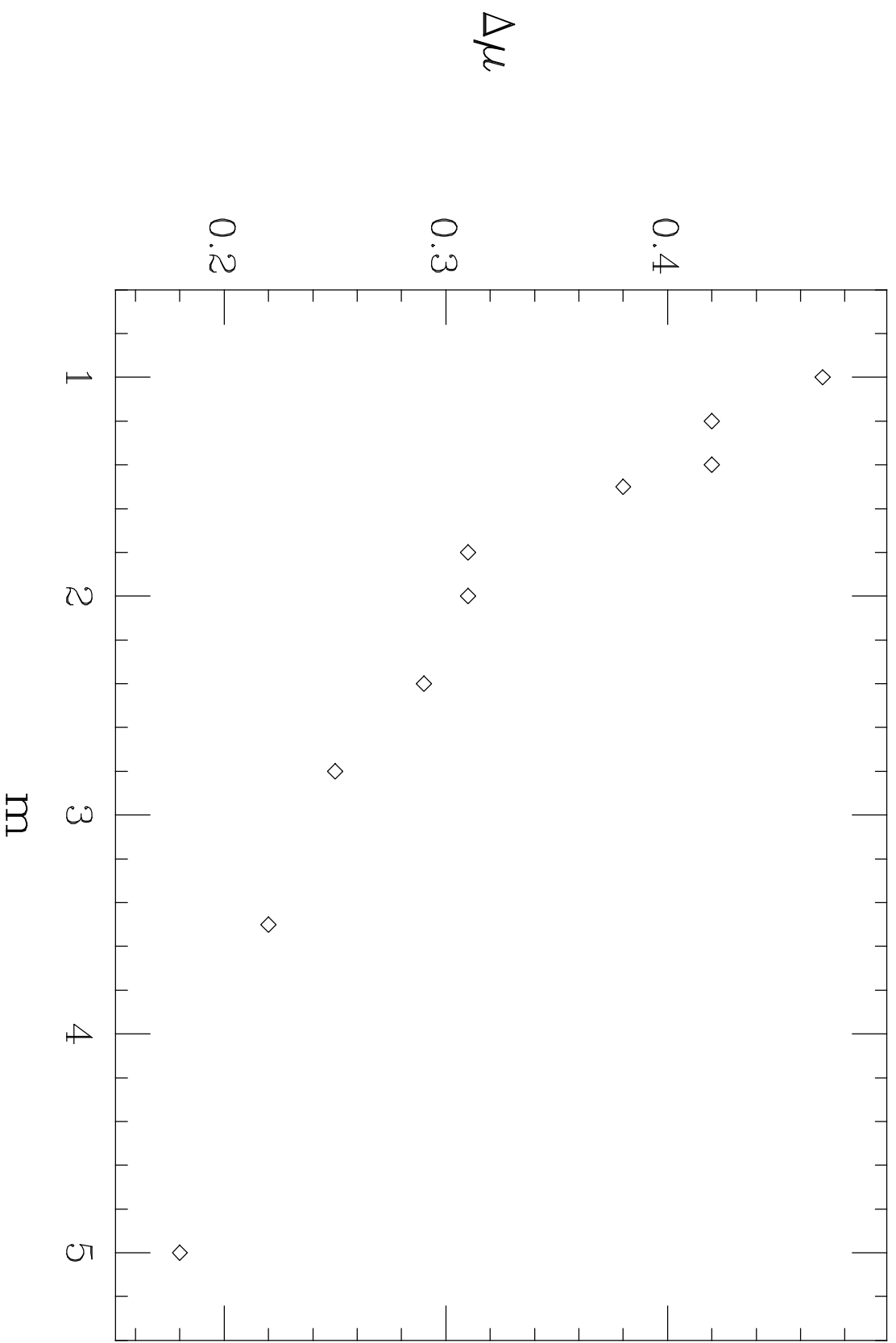
\item{Fig. 4: Width of the $\mu$ region $(\Delta\mu)$
in which the QCD partition function fails to be 
positive versus the quark bare mass $m_q$ in a $4^3\times 4$ lattice at 
$\beta=5.5$.}

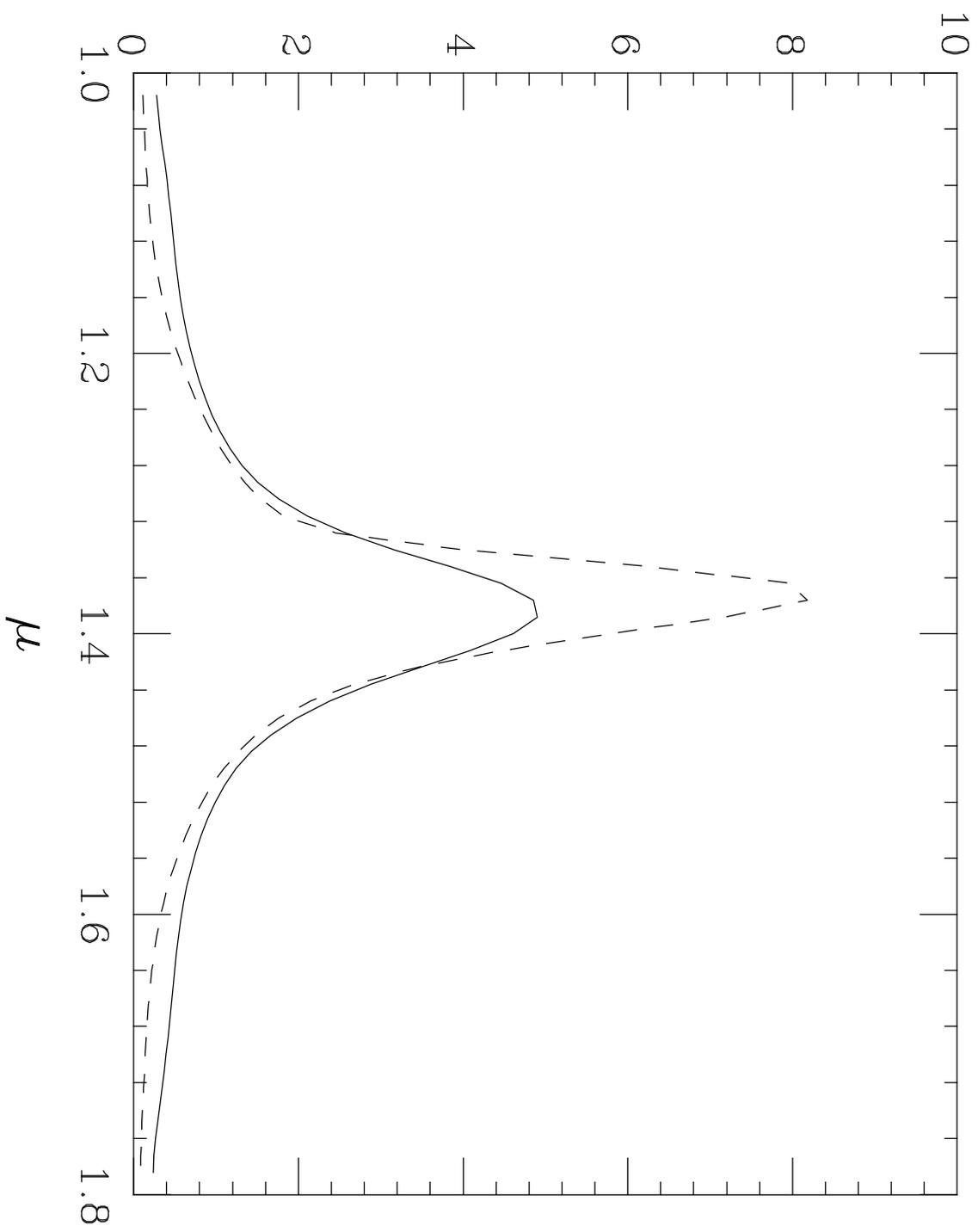
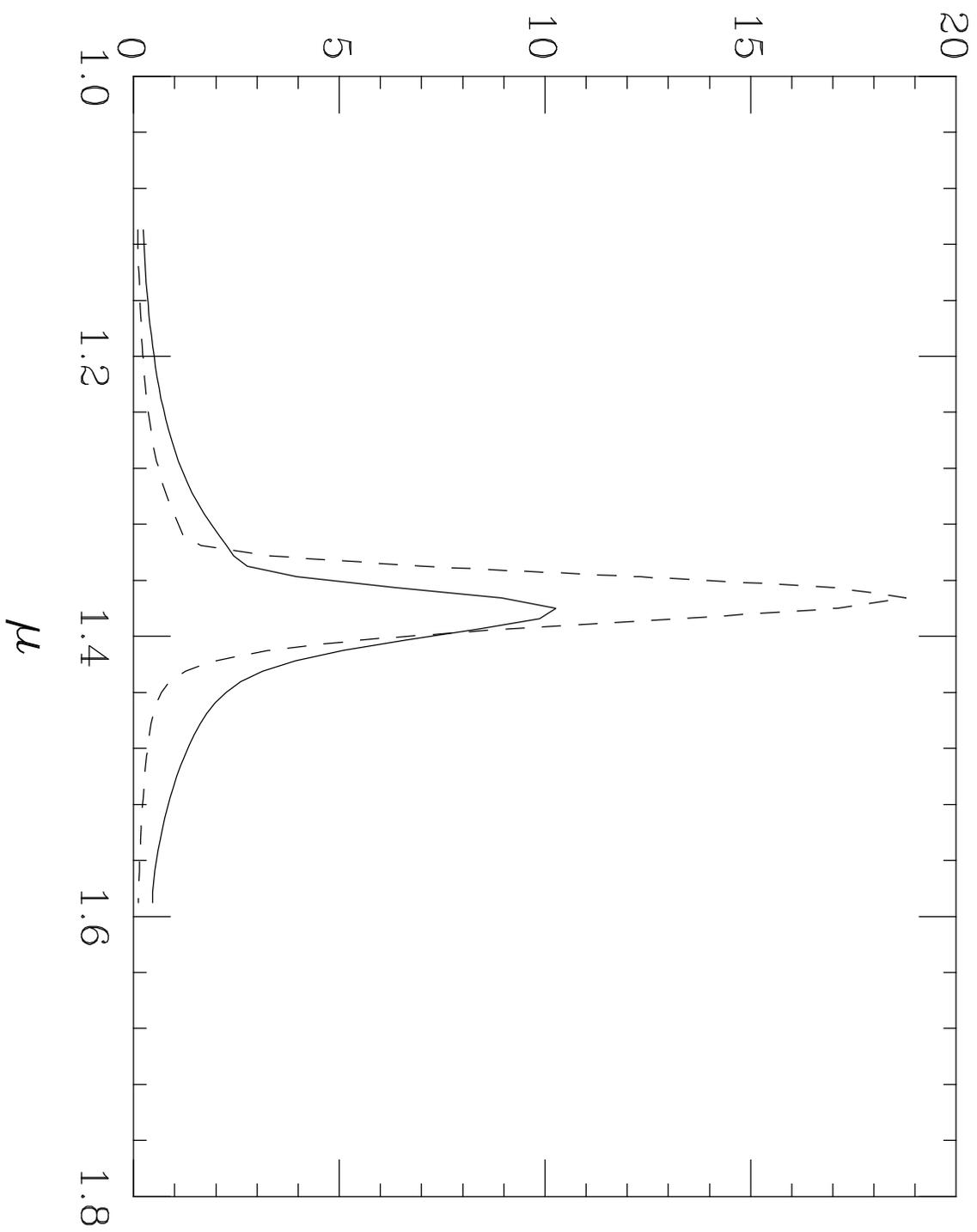
\item{Fig. 5: Derivative of the number density respect to the 
chemical potential in a $4^3\times 4$ (a) and $6^3\times 4$ (b) 
lattice at $m_q=1.5$ for 
$\beta=5.3$ (dashed line), and $\beta=5.7$ (continuous line).}

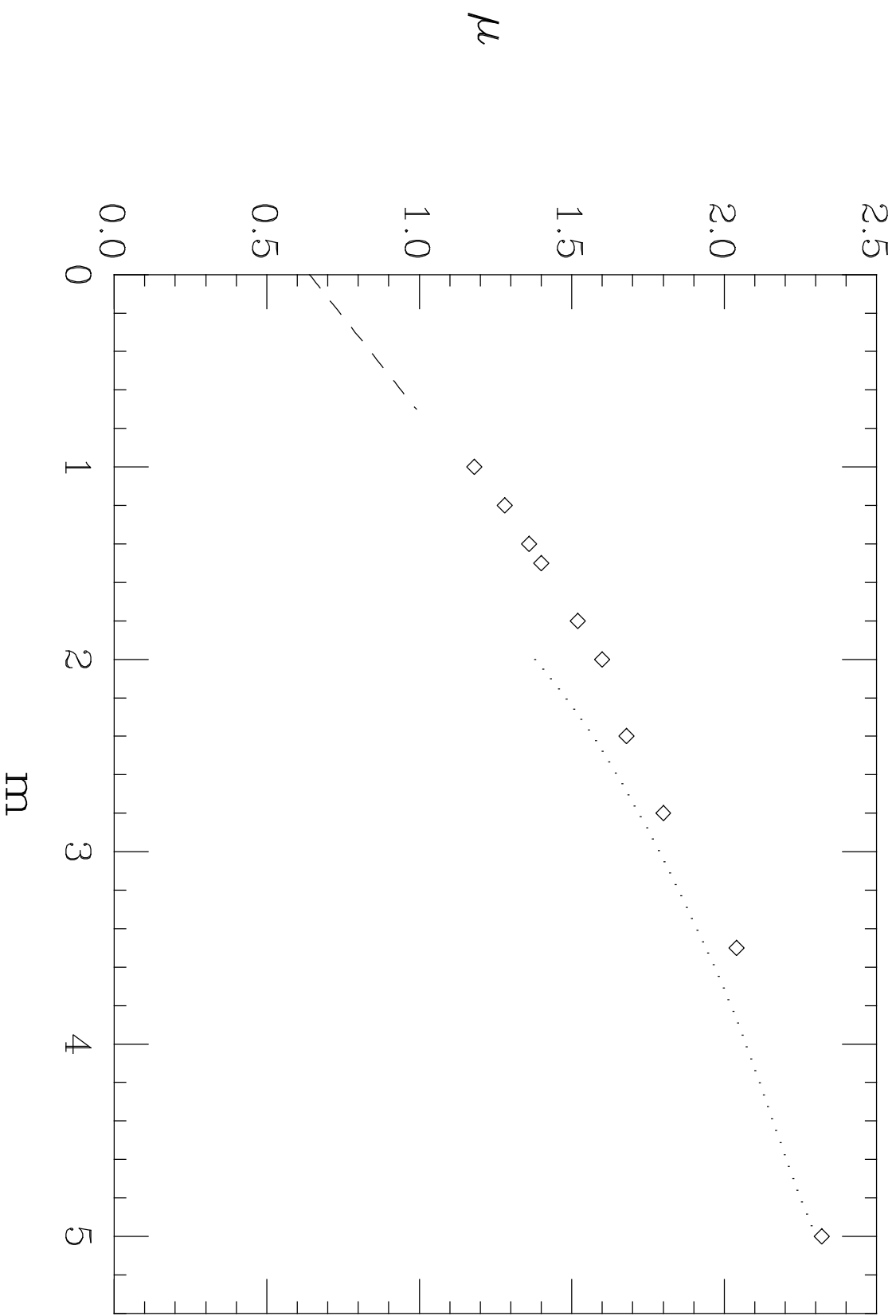
\item{Fig. 6: Saturation critical chemical potential $\mu_c^S$ versus the 
quark bare mass $m_q$ in a $4^3\times 4$ lattice at $\beta=0$. Dotted line 
is the large mass limit, the dashed one is the result of \cite{karsh}.}

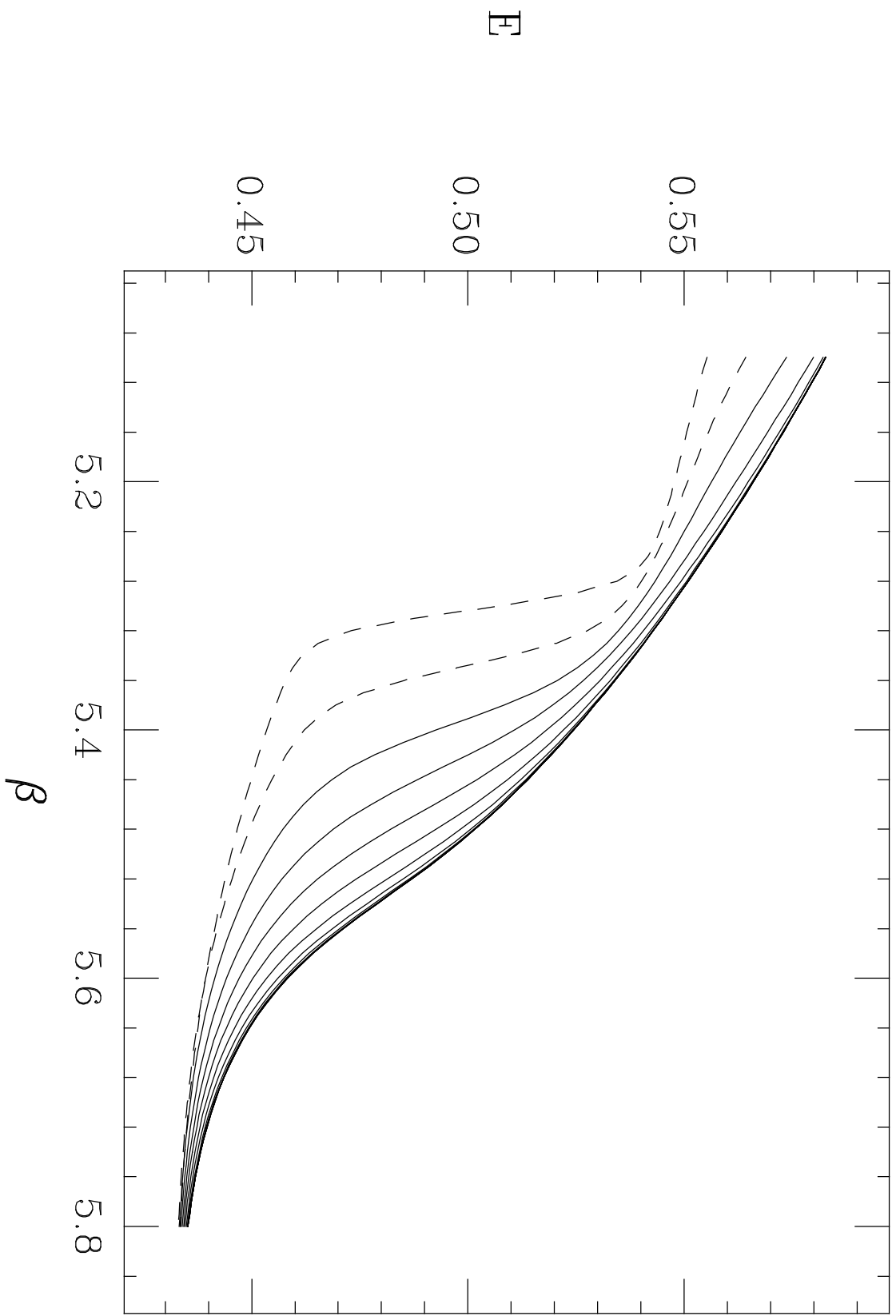
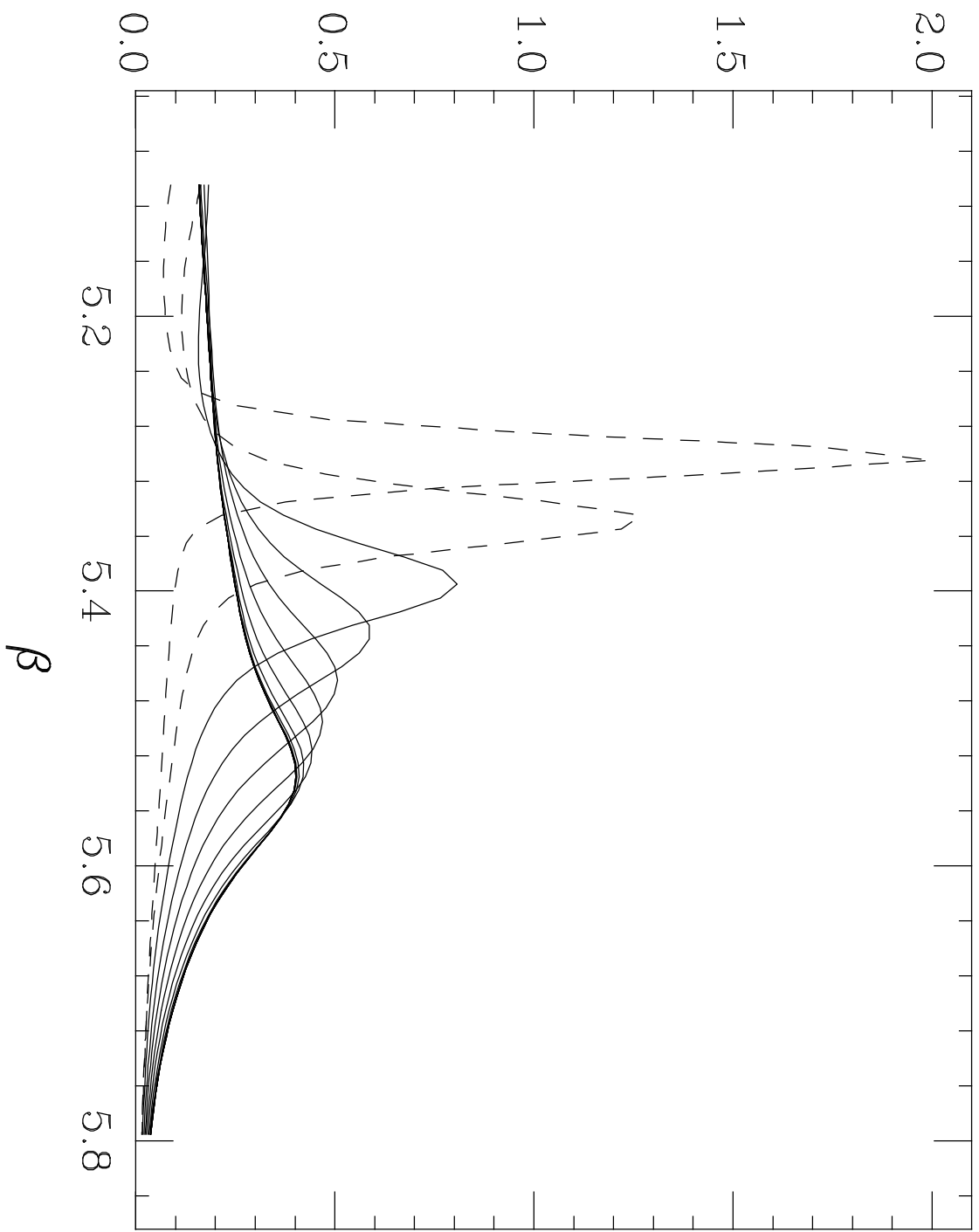
\item{Fig. 7: Plaquette energy $E(\beta,\mu)$ (a) and its derivative
$\partial E(\beta,\mu)/\partial\beta$ (b)
evaluated in a $4^3\times 4$
lattice at $m_q=1.8$ for $\mu=0.0 \to 1.5$ (from the right to the left) 
in steps of $0.1$. Dashed line are for $\mu>\mu_1$.}

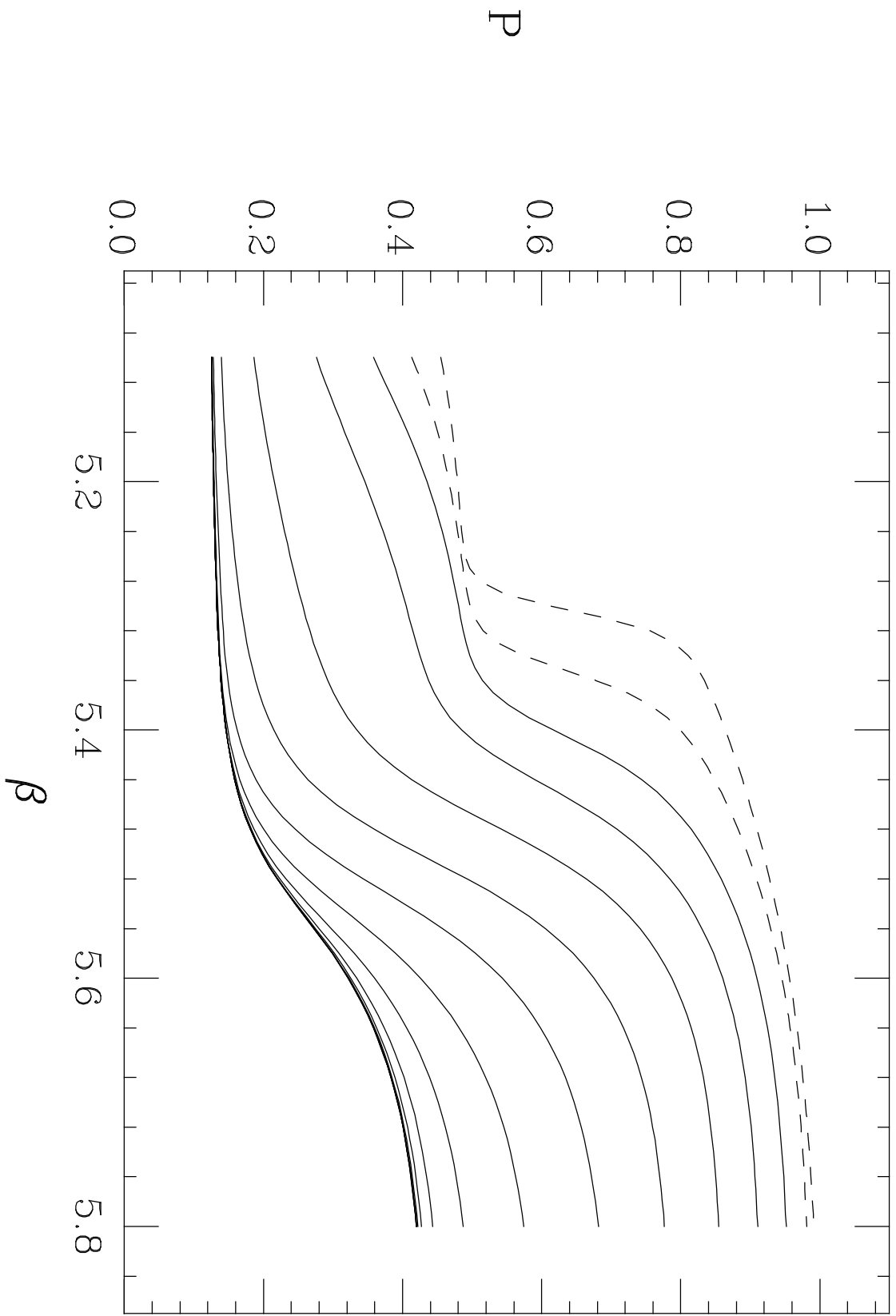
\item{Fig. 8: Polyakov loop $P(\beta,\mu)$ evaluated in a $4^3\times 4$
lattice at $m_q=1.8$ for $\mu=0.0 \to 1.5$ (from the right to the left) 
in steps of $0.1$. Dashed line are for $\mu>\mu_1$. } 

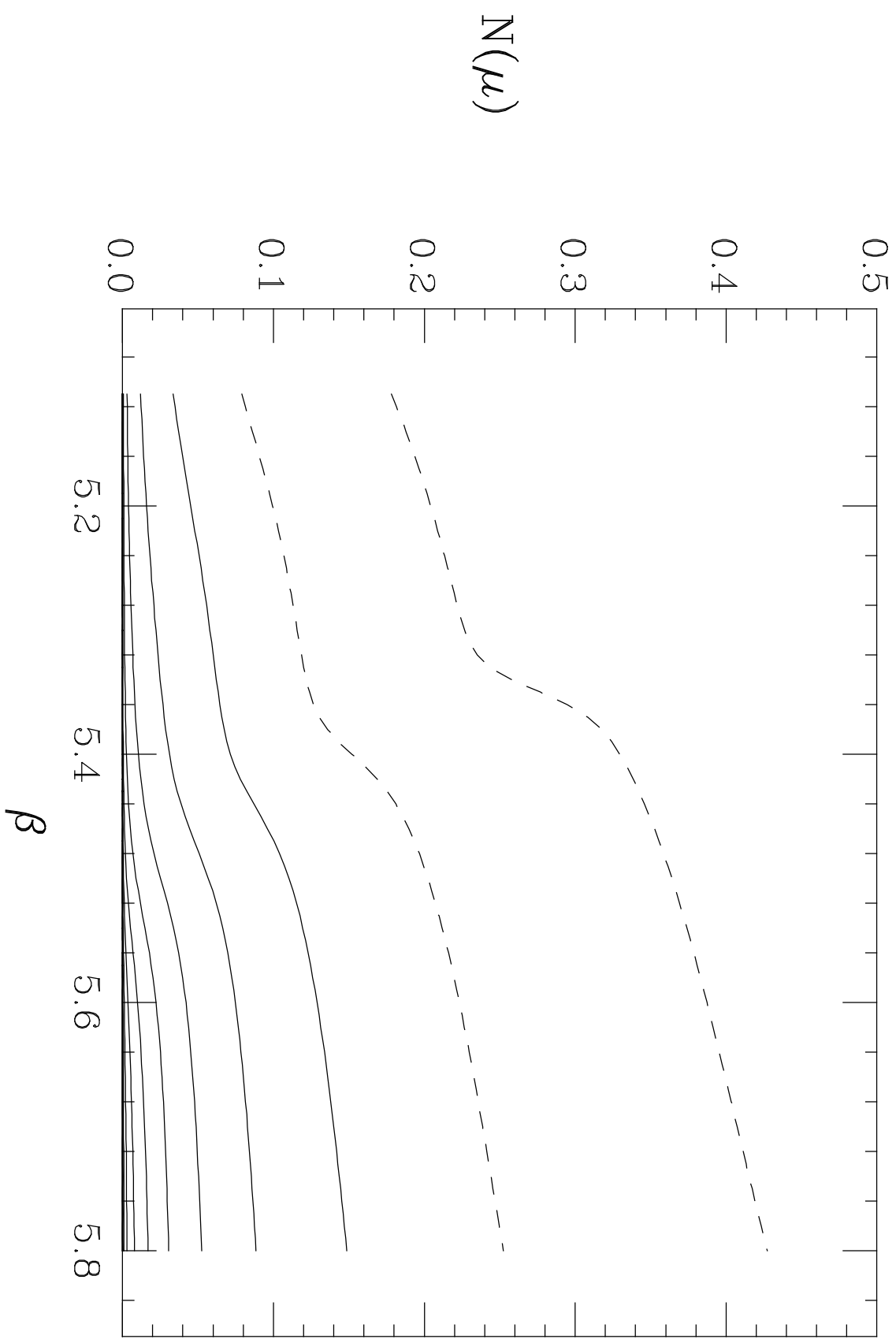
\item{Fig. 9: Number density evaluated in a $4^3\times 4$
lattice at $m_q=1.8$ for $\mu=0.0 \to 1.5$ (from the right to the left) 
in steps of $0.1$. Dashed line are for $\mu>\mu_1$. } 

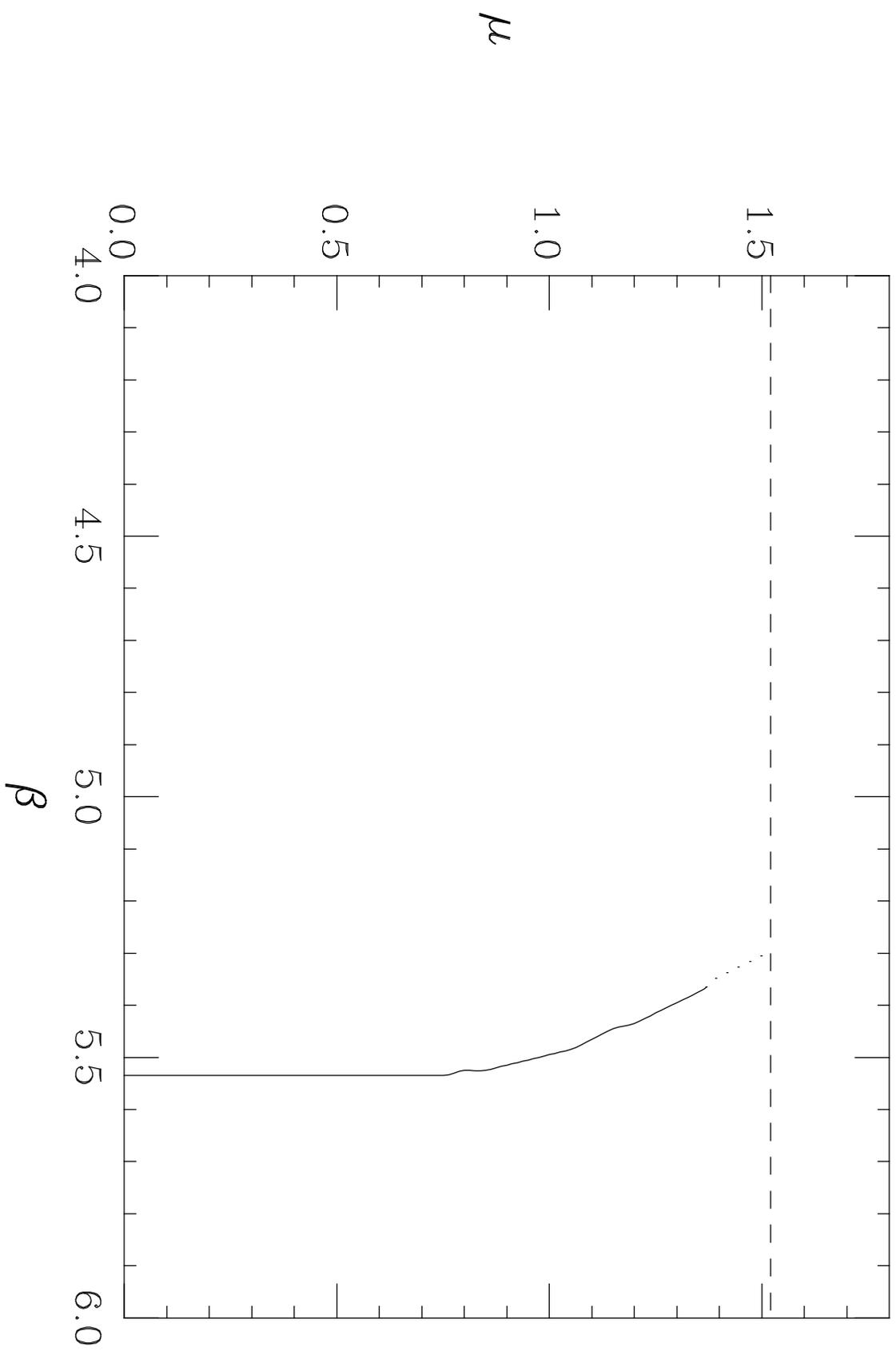
\item{Fig. 10: Phase diagram for the $4^3\times 4$ lattice at
$m_q=1.8$ in the $(\beta,\mu)$ plane, the dotted line is for $\mu>\mu_1$.}

\end{itemize}


\newpage
\vskip 1 truecm
\begin{thebibliography}{9}

\bibitem{bhakar}
G. Bhanot, K. Bitar, R. Salvador, Phys. Lett. {\bf B188} \rm (1987) 246; 
M. Karliner, S. Sharpe, Y.F. Chang, Nucl. Phys. {\bf B302} \rm (1988) 204. 

\bibitem{azc}
V. Azcoiti, I.M. Barbour, R. Burioni, G. Di Carlo, A.F. Grillo, G. Salina, 
Phys. Rev. {\bf D51} \rm (1995) 5199.

\bibitem{noi1}
R. Aloisio, V. Azcoiti, G. Di Carlo, A. Galante, A.F. Grillo,
Phys. Lett. {\bf B428} \rm (1998) 166. 

\bibitem{noi2}
R. Aloisio, V. Azcoiti, G. Di Carlo, A. Galante, A.F. Grillo,
Phys. Lett. {\bf B435} \rm (1998) 175. 

\bibitem{bilic}
N. Bilic, K. Demeterfi,
Phys. Lett. {\bf B212} \rm (1988) 83. 

\bibitem{lat97}
R. Aloisio, V. Azcoiti, G. Di Carlo, A. Galante, A.F. Grillo,
Nucl. Phys. Proc. Suppl. {\bf 63} \rm (1998) 442.

\bibitem{karsh}
F. Karsch, K.H. M\"utter,
Nucl. Phys. {\bf B313} \rm (1989) 541. 

\bibitem{bar}
I.M. Barbour, S.E. Morrison, E.G. Klepfish, J.B. Kogut, M.P. Lombardo,
Phys. Rev. {\bf D56} \rm (1997) 7063.

\bibitem{gibbs}
P.E. Gibbs,
Phys. Lett. {\bf B172} \rm (1986) 53. 

\bibitem{noi3}
R. Aloisio, V. Azcoiti, G. Di Carlo, A. Galante, A.F. Grillo,
hep-lat 9811033.

\bibitem{hase}
J.B. Kogut, H. Matsuoka, M. Stone, H.W. Wyld, S. Shenker,
J. Shigemitsu, D.K. Sinclair, Nucl. Phys. {\bf B225} [FS9] \rm (1983) 93;
P. Hasenfratz, F. Karsch, Phys. Lett. {\bf B125} \rm (1983) 308. 

\bibitem{mfa}
V. Azcoiti, G. Di Carlo, A.F. Grillo, Phys. Rev. Lett. {\bf 65} 
\rm (1990) 2239; V. Azcoiti, A. Cruz, G. Di Carlo, A.F. Grillo and 
A. Vladikas, 
Phys. Rev. {\bf D43} \rm (1991) 3487; 
V. Azcoiti, G. Di Carlo, L.A. Fernandez, A. Galante, A.F. Grillo, V. Laliena, 
X.Q. Luo, C.E. Piedrafita and A. Vladikas, 
Phys. Rev. {\bf D48} \rm (1993) 402; V. Azcoiti, G. Di Carlo, 
A. Galante, A.F. Grillo, V. Laliena, 
Phys. Rev. {\bf D50} \rm (1994) 6994.

\bibitem{ks}
H. Kluberg-Stern, A. Morel, B. Peterson,
Nucl. Phys. {\bf B215} [FS7] \rm (1983) 527.

\end{thebibliography}
\end{document}